\pdfoutput=1
\documentclass[aps,twocolumn,superscriptaddress, prb]{revtex4-2}
\usepackage{graphicx}
\usepackage{dcolumn}
\usepackage{bm}
\usepackage{amsmath}
\usepackage{amssymb}
\usepackage{braket}
\usepackage[caption=false]{subfig}
\usepackage[colorlinks=true]{hyperref}

\usepackage{physics}
\usepackage{mathrsfs}
\usepackage{comment}

\def\*#1{\mathbf{#1}}

\def\t#1{\text{#1}}

\def\tt#1{\textit{#1.}|}

\begin{document}

\title{Noise-Induced Localized Patterns in Excitable Media: Amplitude versus Persistence}

 

\author{Chau Dao}
\affiliation{Department of Physics and Astronomy and Bhaumik Institute for Theoretical Physics, University of California, Los Angeles, California 90095, USA}

\author{Jr-Ming Yang}
\affiliation{Department of Pathology, Department of Cell Biology, Center for Cell Dynamics, and Institute for NanoBioTechnology, Johns Hopkins University School of Medicine, Baltimore, MD 21287, USA}

\author{Chuan-Hsiang Huang}
\affiliation{Department of Pathology, Department of Cell Biology, Center for Cell Dynamics, and Institute for NanoBioTechnology, Johns Hopkins University School of Medicine, Baltimore, MD 21287, USA}

\author{Gia-Wei Chern}
\affiliation{Department of Physics, University of Virginia, Charlottesville, Virginia 22904, USA}

\begin{abstract}
Transient spatially localized activity underlies a broad range of biological processes, yet the statistical property of fluctuations that controls its nucleation remains unclear. We systematically investigate noise-induced dynamics in a spatially extended FitzHugh-Nagumo excitable system and identify three regimes: a quiescent phase, a spatially extended alternating phase, and a pattern-forming phase characterized by transient localized excitation patches. Surprisingly, we find that temporal noise correlations are not required for patch formation: Gaussian white noise produces localized patches when its amplitude is sufficiently large, whereas weaker fluctuations can achieve the same effect when temporal correlations allow them to persist. Our results identify the instantaneous amplitude and the persistence as joint stochastic control parameters for transient pattern formation in excitable media. These distinct quantities are linked through the integrated noise strength, which quantifies the accumulated stochastic forcing available to nucleate an excitation. In addition, the inhibitor response time subsequently determines whether the excitation remains localized or spreads throughout the system. 
\end{abstract}

\maketitle

\begin{figure}
     \centering
     \includegraphics[width=1\linewidth]{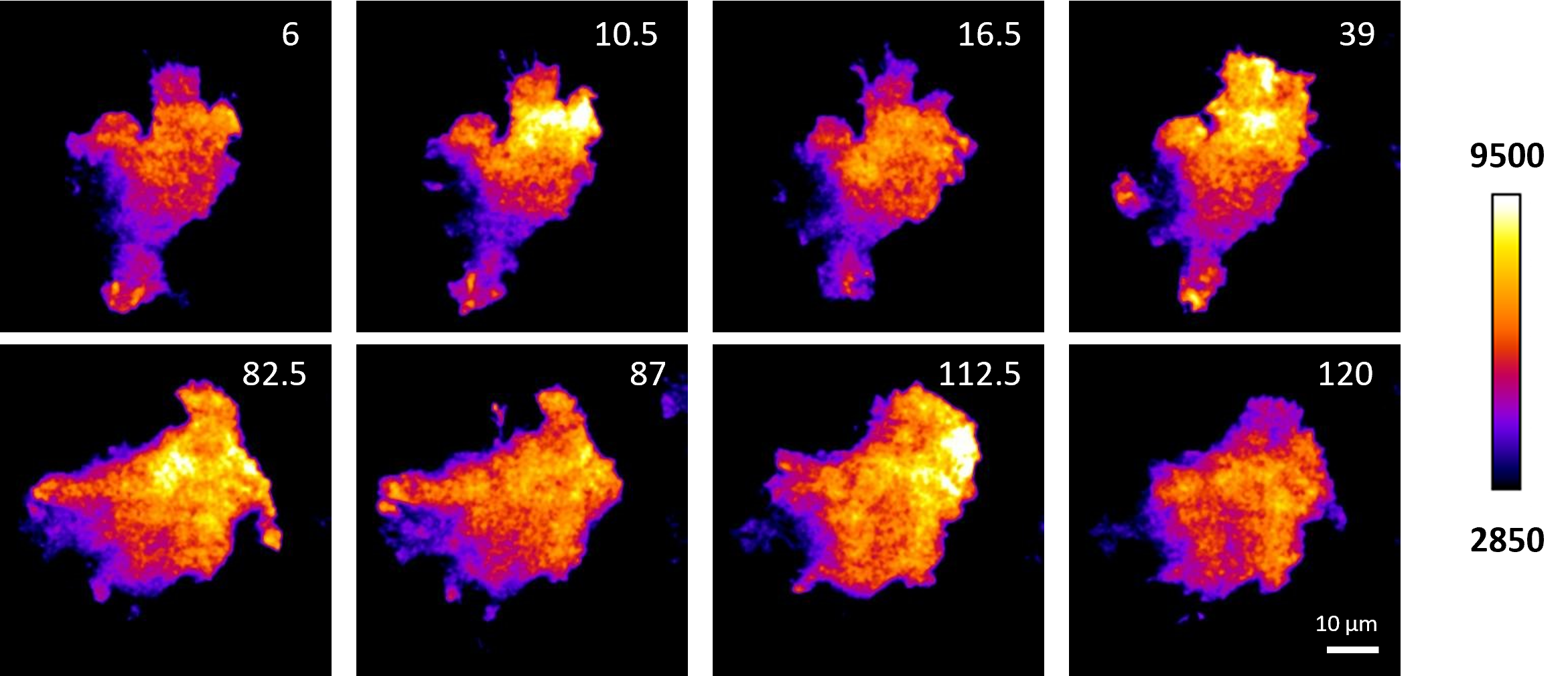}
     \caption{Time-lapse images of a 4T1 tumor cell expressing the biosensor PH-AKT, which detects the activity of phosphoinositide 3-kinase (PI3K)~\cite{watton99}. The images were acquired using total internal reflection fluorescence microscopy, which captures fluorescence signals very close ($< 200$~nm) to the cover slip under the basal cell surface. Time stamp: min. Color scale: fluorescence intensity (arbitrary units).}
     \label{fig:exp}
 \end{figure}

\tt{Introduction}Biological functions emerge from complex networks of dynamically interacting molecules. In eukaryotic cell motility, the molecular networks that coordinate cell movement have been shown to exhibit the characteristic behavior of excitable reaction-diffusion systems~\cite{vicker00,gunther04,weiner07,asano08,huang13,arai10,allard13,taniguchi13,xiong10,nishikawa14,haastert17,devreotes17,matsuoka24}. These networks couple signaling proteins, including Ras and PI3K, to cytoskeletal components such as actin and its regulators. Intrinsic molecular fluctuations can stochastically activate the network, generating a rich variety of patterns, including propagating waves and transient, spatially localized patches of activity. An example is shown in Fig.~\ref{fig:exp}, where PI3K activity at the basal surface of a cell was visualized using the fluorescence biosensor PH-AKT. Remarkably, even in the absence of external stimuli, the cell exhibited continual morphological changes accompanied by transient patches of PI3K activity that appeared spontaneously at different locations on the cell surface.

These spatiotemporal signaling patterns play essential roles in cellular function. Waves and localized patches of Ras-PI3K activity drive the formation of protrusions during cell movement~\cite{asano08,huang13,allard13,haastert17,zhan20,miaoMSB19} and can trigger pulsatile activation of downstream pathways, including the extracellular signal-regulated kinase (ERK) pathway, a key regulator of cell proliferation~\cite{aoki13,albeck13,yang18}. These waves also coordinate the spatiotemporal dynamics of glycolytic enzymes involved in energy production~\cite{zhan25}. Dysregulation of the underlying excitable network can disrupt both cell growth, motility, and metabolism. Moreover, activating mutations affecting the Ras-PI3K signaling network occur in more than 50\% of human cancers~\cite{vega18} and have been associated with increased network excitability~\cite{zhan20,yang18,zhan25}. Hence, understanding the dynamical mechanisms that select among different signaling patterns is crucial for elucidating both normal cellular behavior and its pathological disruption.

\begin{figure*}
    \centering
    \includegraphics[width = \linewidth]{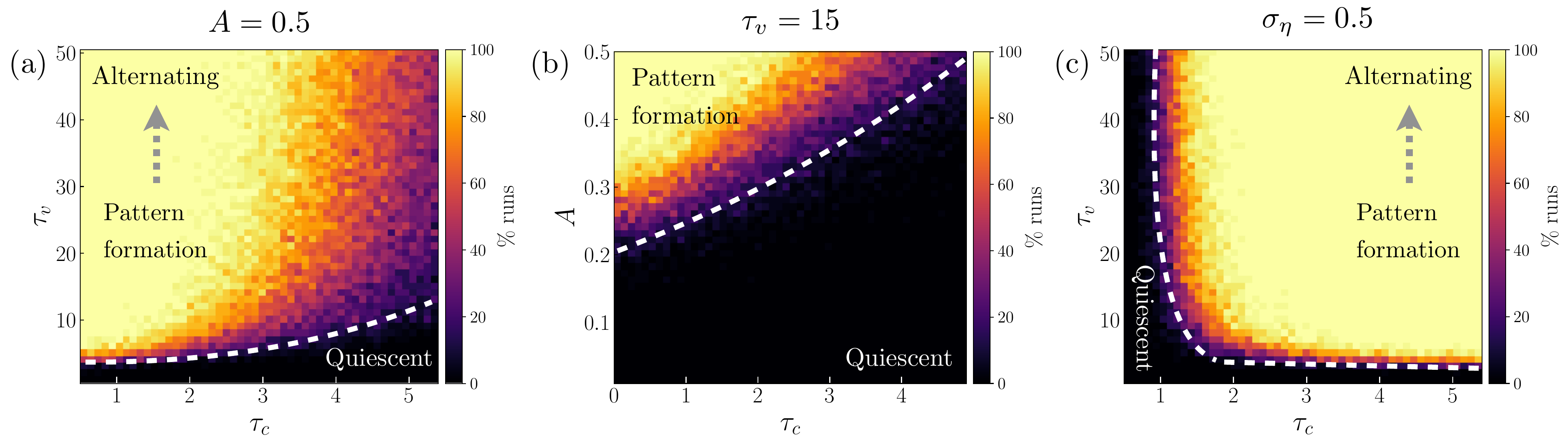}
    \caption{Noise-induced dynamical phases under three parameterizations of the active noise. Color indicates the percentage of 50 independent runs exhibiting noise-induced excitation. Panel (a) is obtained by varying the inhibitor response time $1\leq\tau_v\leq50$ and correlation time $0.5\leq\tau_c<5.5$ at fixed integrated noise strength $A=0.5$. Panel (b) is obtained by sweeping the ranges $0.01\leq A\leq0.5$ and $0\leq\tau_c\leq4.9$ at fixed $\tau_v=15$. Panel (c) is obtained by varying $1\leq\tau_v\leq50$ and $0.5\leq\tau_c<5.5$ at fixed equal-time rms noise amplitude $\sigma_\eta=0.5$. The white dashed curves delineate the boundary of the quiescent regime, while the arrows in (a) and (c) indicate the crossover from localized pattern formation to the spatially extended alternating phase. The remaining parameters are $\alpha=0.5$, $\beta=0.7$, $D_u=1$, and $D_v=5$.}
    \label{fig:phase-diagram}
\end{figure*}
 
These observations raise a broader question: what properties of stochastic fluctuations determine whether an excitable medium remains quiescent, develops system-wide activity, or sustains production of transient localized patterns? In particular, is nucleation controlled by the instantaneous noise amplitude, the duration over which a fluctuation persists, or by the total stochastic forcing accumulated over time? To address these questions, we consider the FitzHugh-Nagumo (FHN) model. Originally introduced as a reduced description of excitation and recovery in nerve membranes~\cite{fitzhugh61,nagumo62} and now widely used as a paradigmatic model of excitable dynamics, the model couples a rapidly responding activator to a slower inhibitory recovery variable. Once the activator crosses an excitation threshold, nonlinear amplification drives it toward an excited state, after which the inhibitor terminates the excitation and restores the resting state. Promoting these variables to spatially varying fields and introducing diffusive coupling yields a minimal reaction-diffusion description of an excitable medium capable of supporting propagating waves and other spatially organized structures~\cite{nagumo62,rinzel73,buschPRE03}. Intrinsic and environmental fluctuations can then be incorporated through a stochastic Langevin force, conventionally represented by Gaussian white noise~\cite{pikovskyPRL97}.

Although white noise is mathematically convenient, fluctuations in physical, chemical, and biological systems often develop over finite times and hence possess nontrivial temporal correlations~\cite{hanggi95,kaern05,raser05,briat23}. Such persistent fluctuations arise naturally in nonequilibrium and active systems and are commonly modeled using colored noise~\cite{ramaswamy10,marchetti13,elgeti15,cates15,bechinger16,martin2021,tevrugt26}. Temporal correlations have been shown to strongly influence noise-induced dynamics in FitzHugh--Nagumo and other excitable systems~\cite{brugioni05,brandstetter10,buschPRE03,gu18}, as well as wave nucleation and pattern formation in spatially extended reaction--diffusion systems~\cite{beato05,balanov06,perc05,wang06,das13,adamer20}. Finite-duration forcing has also been invoked to generate transient localized structures, partly because strong white noise was expected to obscure coherent spatial patterns~\cite{hecht2010,hechtPCB11}.

Here, we systematically interpolate between white and temporally correlated Gaussian noise to identify the key drivers for pattern formation.
We find that transient localized patches can be nucleated either by short-lived fluctuations with sufficiently large instantaneous amplitude or by weaker fluctuations that persist for longer durations. These two routes are unified by the integrated noise strength, which measures the stochastic forcing accumulated over time. The inhibitor response time then determines whether a nucleated excitation is suppressed locally or spreads into system-wide alternating activity. The instantaneous amplitude as well as the temporal persistence emerge on equal footing as central stochastic quantities governing the formation of transient localized patterns.

\tt{Reaction-diffusion equations}To isolate the essential ingredients underlying stochastic pattern formation, we consider a minimal one-dimensional FitzHugh-Nagumo reaction-diffusion system defined on a circle. We denote the spatial coordinate by $\ell$ and impose periodic boundary conditions. The activator and inhibitor fields, $u(\ell,t)$ and $v(\ell,t)$, evolve according to
\begin{subequations}
\label{eq:fhn}
    \begin{align}
        \frac{\partial u}{\partial t}
        &=D_u\partial_\ell^2u
        +\frac{1}{\tau_u}(1-u^2)(u-v)
        +\eta(\ell,t),
        \label{eq:fhn_u}\\
        \frac{\partial v}{\partial t}
        &=D_v\partial_\ell^2v
        +\frac{1}{\tau_v}(u-\alpha v+\beta).
        \label{eq:fhn_v}
    \end{align}
\end{subequations}
Here, $D_u$ and $D_v$ are the diffusion coefficients of the activator and inhibitor, while $\tau_u$ and $\tau_v$ set their characteristic response times. The parameters $\alpha$ and $\beta$ determine the relaxation and homogeneous steady-state value of the inhibitor. We focus on the regime $D_v>D_u$, in which the inhibitor diffuses more rapidly than the activator, and take $\alpha<\beta$. In the absence of noise, the system has a quiescent homogeneous state given by $u_0=-1$ and $v_0=(\beta-1)/\alpha$. Stochastic fluctuations enter Eq.~\eqref{eq:fhn_u} through the additive noise field $\eta(\ell,t)$ acting on the activator.

\tt{Ornstein-Uhlenbeck active noise}To investigate how temporal persistence affects the dynamics of the excitable medium, we model $\eta(\ell,t)$ as an Ornstein-Uhlenbeck process with zero mean and autocovariance
\begin{equation}
\label{eq:noiseproperties}
\left\langle
\eta(\ell,t)\eta(\ell',t')
\right\rangle
=
\delta(\ell-\ell')\,
\sigma_\eta^2
e^{-|t-t'|/\tau_c},
\end{equation}
where $\sigma_\eta$ denotes the equal-time root mean square (rms) amplitude of the stochastic forcing and $\tau_c$ is the temporal correlation time. An important quantity, which will emerge as a key quantity to unify the pattern-formation mechanisms is the \emph{integrated noise strength}
\begin{align}
\label{eq:integratednoise}
A \equiv \int_0^\infty \t dt\, \left\langle \eta(\ell,0)\eta(\ell,t) \right\rangle = \sigma_\eta^2\tau_c,
\end{align}
which measures the total stochastic forcing accumulated over the correlation time. Henceforth, we regard $A$, rather than $\sigma_\eta$, as the primary control parameter.

Equivalently, the noise field evolves according to
\begin{equation}
\label{eq:noiseeq}
    \frac{\partial\eta(\ell,t)}{\partial t}
    =-\frac{\eta(\ell,t)}{\tau_c}
    +\frac{\sqrt{2A}}{\tau_c}\,\xi(\ell,t),
\end{equation}
where $\xi(\ell,t)$ is a Gaussian white noise field satisfying
$\langle\xi(\ell,t)\rangle=0$ and
$\langle\xi(\ell,t)\xi(\ell',t')\rangle
=\delta(\ell-\ell')\delta(t-t')$.
To simulate the nonlinear FitzHugh-Nagumo reaction-diffusion equations with active noise, we employ the Crank-Nicolson method; details of the numerical implementation are presented in Appendix~\ref{sec:cranknicolson}.

\begin{figure}
    \centering
    \includegraphics[width = \linewidth]{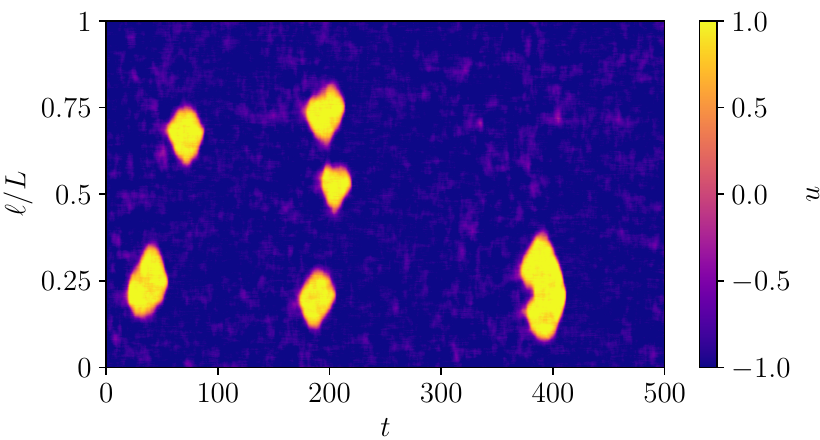}
    \caption{Spacetime diagram of the activator field $u(\ell,t)$, illustrating the spontaneous emergence of transient, spatially localized excitation patches. Time is plotted horizontally, while the vertical axis gives the normalized length $\ell/L$ along the periodic domain; the color scale indicates the local activator value. Starting from the quiescent background $u\simeq-1$, localized excited regions ($u \simeq 1$) nucleate at different positions and times, evolve over finite lifetimes, and eventually decay. The parameters are $\alpha=0.5$, $\beta=0.7$, $\tau_v/\tau_u=25$, $\tau_c/\tau_u=5$, $\Delta t/\tau_u=0.1$, and $A=1.25$.}
    \label{fig:simulation}
\end{figure}

\tt{Dynamical phases}Figure~\ref{fig:phase-diagram} summarizes the noise-induced dynamics of the stochastic FHN system. Figs.~\ref{fig:phase-diagram}(a) and (c) show phase diagrams obtained by varying the inhibitor response time $\tau_v$ and the noise correlation time $\tau_c$ at fixed integrated noise strength $A$ and fixed equal-time rms amplitude $\sigma_\eta$, respectively. Figure~\ref{fig:phase-diagram}(b) shows the corresponding dependence on $A$ and $\tau_c$ at fixed $\tau_v$. Each point represents the cumulative data from 50 runs, each of duration $500\tau_u$. Together, these complementary parameterizations disentangle the roles of inhibitor dynamics, noise amplitude, and temporal persistence in controlling the emergence of localized excitation patches.
Three dynamical regimes emerge: a quiescent phase, a spatially extended alternating phase, and an intermediate pattern-forming phase characterized by transient, spatially localized excitation patches. We now discuss the physical mechanisms underlying each regime.

In the quiescent phase, the activator remains near its homogeneous steady-state value $u_0=-1$, apart from small fluctuations. For $-1<u<1$, the local reaction term in Eq.~\eqref{eq:fhn_u} amplifies the activator only when $u>v$, which near equilibrium corresponds approximately to crossing the threshold $v_0=(\beta-1)/\alpha$. The system remains quiescent when fluctuations are either too weak or too short-lived to overcome this threshold and initiate nonlinear amplification. Quiescence also occurs when the inhibitor responds on a timescale comparable to or shorter than that of the activator, $\tau_v/\tau_u\lesssim1$, because the inhibitory feedback suppresses an incipient excitation before it can develop.

\begin{figure}
    \centering
    \includegraphics[width=\linewidth]{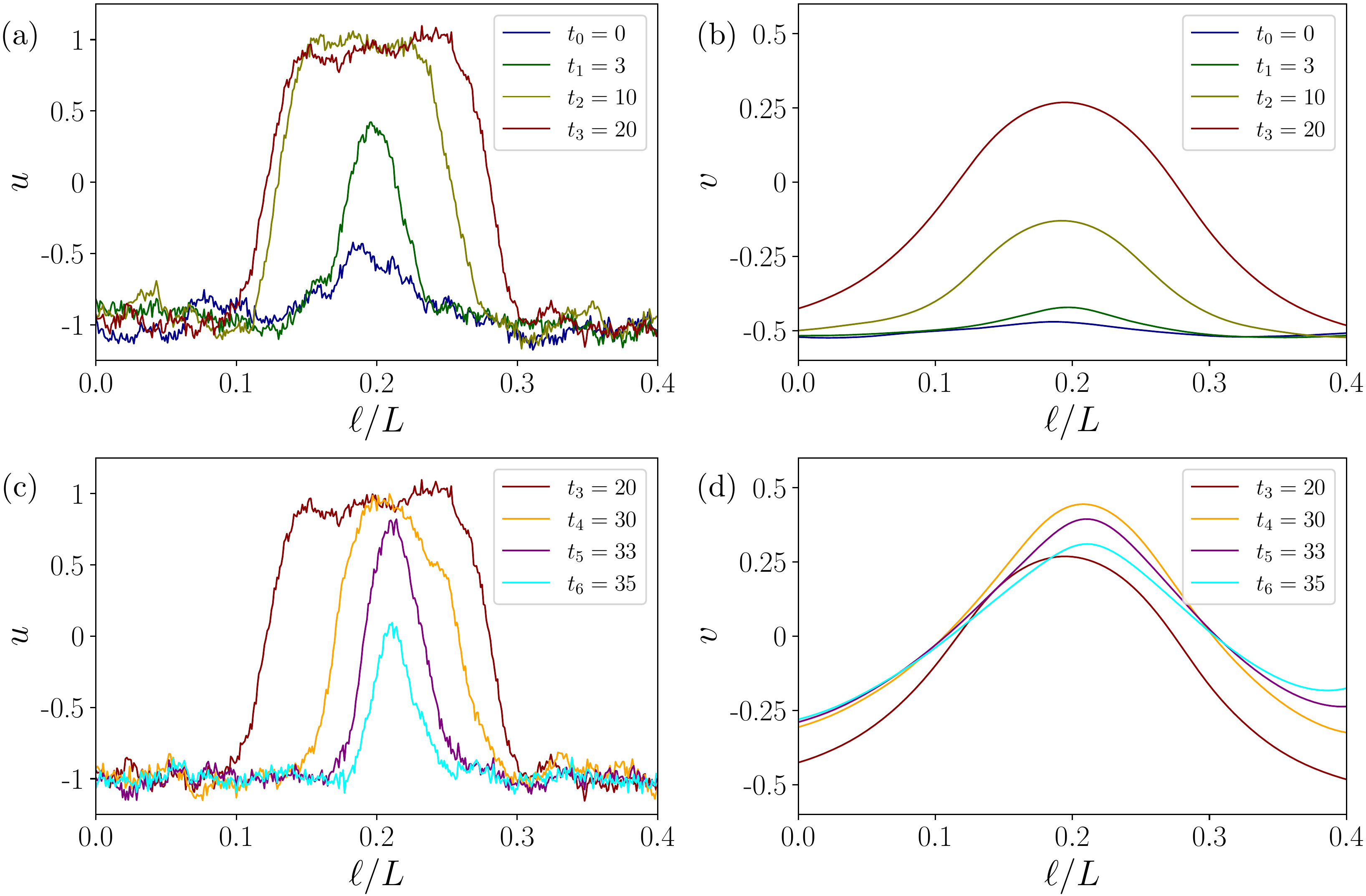}
\caption{Spatial profiles of the activator \(u\) and inhibitor \(v\) over the lifetime of a single excitation patch. Panels (a) and (b) show its noise-induced formation and rapid growth, while panels (c) and (d) show its contraction and decay due to the delayed inhibitor response. Each curve corresponds to a specified time offset.}
    \label{fig:u_and_v}
\end{figure}

At the opposite extreme, a sufficiently slow inhibitor produces a spatially extended alternating phase. When $\tau_v/\tau_u\gg1$, the inhibitor cannot arrest the growth of an excitation before it spreads throughout the system. The activator consequently approaches an approximately uniform excited state with $u\simeq1$. As the inhibitor gradually accumulates, it suppresses the activity and restores the quiescent state, after which the cycle repeats once the inhibitor relaxes. The oscillation period is therefore set primarily by the inhibitor response time $\tau_v$.

The pattern-forming phase lies between these two limits and is the principal focus of this work. Figure~\ref{fig:simulation} shows the spatiotemporal evolution of the activator field $u(\ell,t)$ in the pattern-forming regime. Localized regions with $u\simeq1$ nucleate spontaneously at random positions against the quiescent background, persist for a finite time, and subsequently disappear, after which new patches may emerge elsewhere. Individual nucleation events exhibit similar spatial extent and lifetime, although larger patches occasionally arise through the merger of neighboring excitations.

The activator and inhibitor profiles in Fig.~\ref{fig:u_and_v} illustrate the evolution of an individual patch. A sufficiently strong fluctuation first drives the activator above the local excitation threshold, after which the nonlinear reaction term rapidly amplifies the excitation. Because the inhibitor responds more slowly, the patch is able to grow for a finite interval before inhibitory feedback accumulates, raises the excitation threshold, and terminates the activity. Localized pattern formation therefore requires a separation of timescales: the inhibitor must respond slowly enough to permit nucleation and amplification, yet rapidly enough to arrest the excitation before it becomes system-wide.

Figure~\ref{fig:phase-diagram}(a) shows how the inhibitor dynamics interact with the stochastic forcing at fixed integrated noise strength $A$. Since $\sigma_\eta^2=A/\tau_c$, increasing $\tau_c$ has two competing effects: it increases the temporal persistence of individual fluctuations while simultaneously reducing their equal-time rms amplitude. Temporal persistence favors nucleation by allowing a fluctuation to act coherently over a longer interval, whereas the reduced instantaneous amplitude makes it less likely for the activator to cross the excitation threshold. At fixed $A$, the latter effect eventually dominates as $\tau_c$ increases. Consequently, sustaining localized pattern formation at larger $\tau_c$ requires a slower inhibitor response, producing the upward-sloping quiescent--pattern boundary in Fig.~\ref{fig:phase-diagram}(a).

Figure~\ref{fig:phase-diagram}(b) further demonstrates how the integrated noise strength $A$ and correlation time $\tau_c$ jointly control excitation nucleation. For fixed $\tau_v$ and $\tau_c$, increasing $A$ drives the system from the quiescent phase into the localized pattern-forming phase. The critical value of $A$ increases monotonically with $\tau_c$, consistent with the same competition seen in Fig.~\ref{fig:phase-diagram}(a): at fixed $A$, longer-lived fluctuations have a smaller equal-time rms amplitude and therefore require a larger integrated noise strength to achieve nucleation. Once an excitation is nucleated, however, its subsequent evolution depends only weakly on $A$, indicating that its growth and decay are governed primarily by the intrinsic reaction-diffusion dynamics and, in particular, by the inhibitor response time.

Importantly, Fig.~\ref{fig:phase-diagram}(b) also shows that temporal correlations are not required for localized pattern formation. The pattern-forming regime extends continuously to the white noise limit $\tau_c=0$, demonstrating that Gaussian white noise alone can nucleate transient excitation patches when its strength is sufficiently large. In this regime, Gaussian white noise shifts the mean values of $u$ and $v$ away from their deterministic quiescent values, producing a nonequilibrium quiescent background. Fluctuations about this background must drive $u$ above the corresponding nonequilibrium inhibitor level before nonlinear amplification takes over. Once this occurs, localized excitations remain robust in the presence of the white noise fluctuations. The nonequilibrium quiescent background can be characterized analytically within a mean-field treatment; see Appendix~\ref{sec:noneq} for details.

The complementary parameterization in Fig.~\ref{fig:phase-diagram}(c) provides striking direct evidence for the role of temporal persistence. Here, the equal-time rms amplitude is held fixed at $\sigma_\eta=0.5$, so increasing $\tau_c$ increases the duration of the fluctuations without reducing their instantaneous amplitude. The resulting behavior is qualitatively opposite to that in Fig.~\ref{fig:phase-diagram}(a). Increasing $\tau_c$ now strongly promotes localized pattern formation, and the quiescent region rapidly contracts as the fluctuations become more persistent. Thus, fluctuations that are insufficient to nucleate an excitation when short-lived can cross the effective excitation threshold when they persist for a sufficiently long time.

The contrasting trends in Figs.~\ref{fig:phase-diagram}(a) and (c) distinguish the effects of noise amplitude and persistence. At fixed $A$, increasing $\tau_c$ necessarily reduces $\sigma_\eta$ and ultimately suppresses nucleation; at fixed $\sigma_\eta$, increasing $\tau_c$ enhances persistence without this amplitude penalty and instead promotes nucleation. Together, these results show that temporal correlation is neither intrinsically detrimental nor essential to pattern formation. Rather, its effect depends on how the noise is parameterized: both the instantaneous amplitude of a fluctuation and the time over which it persists determine whether sufficient stochastic forcing accumulates to trigger nonlinear excitation.

We emphasize, however, that the fixed-$\sigma_\eta$ parameterization of Fig.~\ref{fig:phase-diagram}(c) does not provide a continuous interpolation to Gaussian white noise. Since $A=\sigma_\eta^2\tau_c$, the integrated noise strength vanishes as $\tau_c\rightarrow0$ at fixed $\sigma_\eta$. The white-noise limit with finite stochastic forcing must instead be taken at fixed $A$, for which $\sigma_\eta^2=A/\tau_c$ diverges as $\tau_c\rightarrow0$. Therefore, quantitative comparisons between colored and Gaussian white noise should be made using Figs.~\ref{fig:phase-diagram}(a) and (b), whereas Fig.~\ref{fig:phase-diagram}(c) isolates the physical effect of increasing temporal persistence at fixed instantaneous noise amplitude.

\begin{figure}
    \centering
    \includegraphics[width=\linewidth]{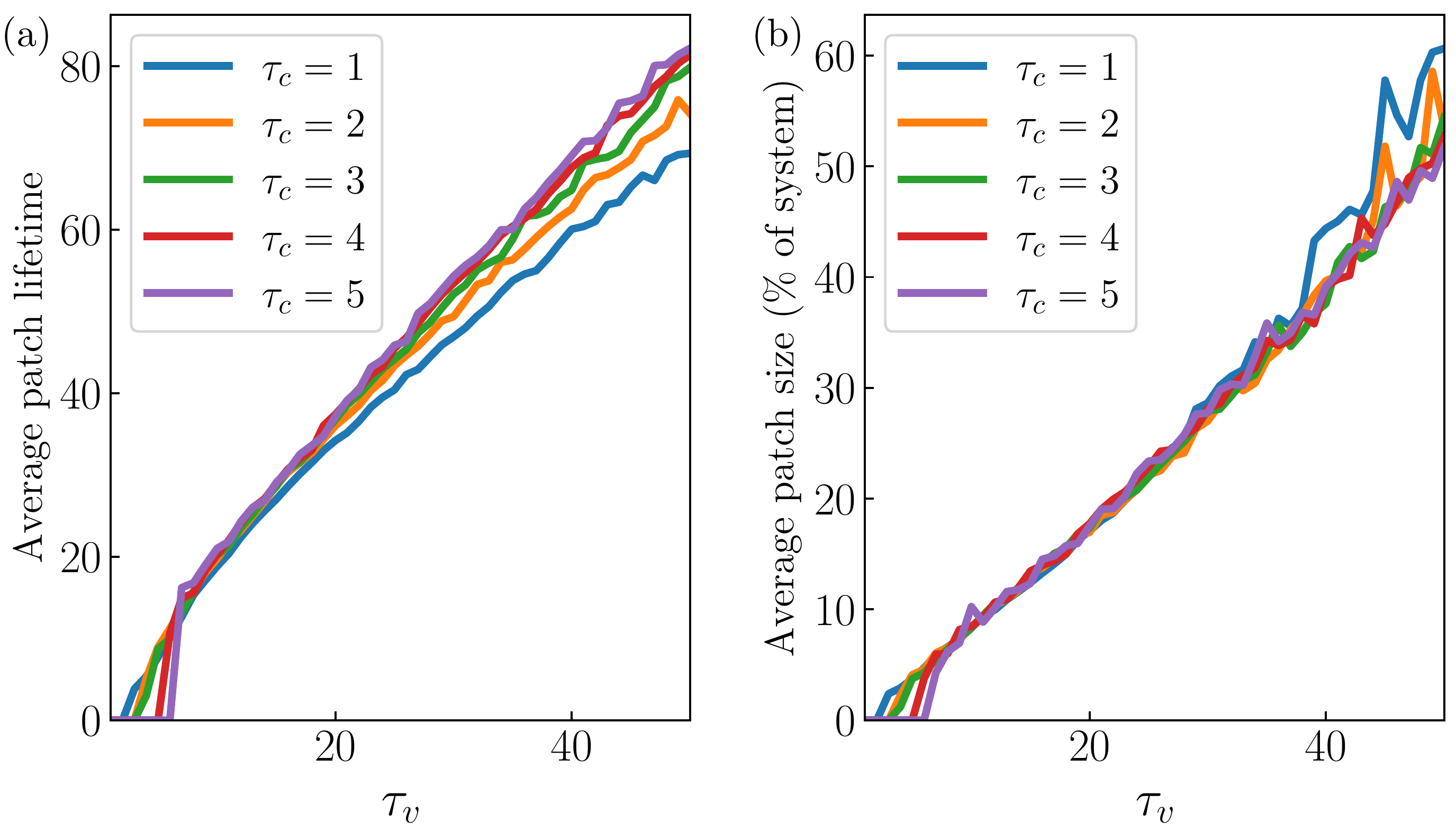}
    \caption{(a) Patch lifetime  (in units of $\tau_u$) and (b) patch size (percentage of system size) as functions of $\tau_v$ for fixed values of $\tau_c = 1,2,3,4,5$. The plots indicate that both patch lifetime and size scale linearly with $\tau_v$.}
    \label{fig:size-lifetime}
\end{figure}

The characteristic spatial extent of a patch can be estimated from straightforward dimensional analysis. Once the patch is formed, the inhibitor rises to extinguish it on the timescale of $\tau_v$. During this interval, the activator front propagates spatially at a speed of order $\sqrt{D_u/\tau_u}$ while the inhibitor diffuses at a speed $\sqrt{D_v/\tau_v}$. The characteristic length scales for the activator and inhibitor are the products of the propagation speeds and $\tau_v$, given by
\begin{equation}
\label{eq:patch-lengths}
\ell_u\sim\tau_v\sqrt{ {D_u}/{\tau_u}}\quad\text{and}
\quad
\ell_v\sim\sqrt{D_v\tau_v},
\end{equation}
respectively. Here, $\ell_u$ estimates the distance traveled by the activator front before inhibitory suppression, whereas $\ell_v$ characterizes the spatial spread of the inhibitor. These estimates are consistent with the profiles in Fig.~\ref{fig:u_and_v}. Since the nonlinear growth of the activator occurs rapidly, $\tau_v$ dominates as the characteristic timescale. See Appendix \ref{section:properties} for more details about the properties of the patches over the ranges swept by the phase diagram in Fig.~\ref{fig:phase-diagram}. 

\tt{Discussion}In this work, we investigated noise-induced pattern formation in a stochastic FitzHugh-Nagumo reaction-diffusion system driven by temporally correlated fluctuations. We identified three dynamical regimes, the most intriguing of which being a pattern-forming phase consisting of transient localized excitation patches. Our central results illustrate that localized patches are nucleated either by strong, short-lived fluctuations or by weaker fluctuations that persist for longer times. The persistence of the fluctuations and the instantaneous fluctuation amplitude emerge on even footing as critical stochastic control parameters governing excitation nucleation. Once nucleated, the subsequent growth, lifetime, and eventual suppression of an excitation are governed predominantly by the inhibitor response time.

These findings provide a simple physical framework for understanding stochastic pattern formation in excitable biochemical signaling networks. In particular, the results suggest that the morphology of transient Ras-PI3K signaling patches depend jointly on the amplitude as well as persistence of intracellular biochemical fluctuations, along with the characteristic recovery time of the inhibitory network. This perspective may help explain how cells exhibiting different sources of molecular fluctuations can nevertheless display similar protrusive dynamics. More broadly, the phase diagrams presented here provide experimentally testable predictions for how changes in signaling noise or inhibitor kinetics regulate the transition between quiescent, localized, and system-wide activity. A potential application of these findings in cell manipulation or therapy is that relatively weak but persistent perturbations may be used to drive specific excitable systems across the excitation threshold, potentially reducing unwanted effects on other systems associated with stronger stimulation.

Beyond cell signaling, we expect our results to apply more broadly to stochastic excitable media in which fluctuations possess finite temporal correlations. Examples include intracellular biochemical networks, active matter systems, ecological and epidemiological reaction-diffusion models, and other nonlinear systems exhibiting threshold dynamics. It would also be interesting to examine whether this principle extends to higher-dimensional excitable media, more realistic signaling networks, and active systems with non-Gaussian or spatially correlated fluctuations.

\begin{acknowledgments}
J.-M.Y. and C.-H.H. were partially supported by the National Institutes of Health through NIGMS Grants No.~R01GM136711 and No.~R35GM163951. G.-W.C. was partially supported by the U.S. Department of Energy, Office of Basic Energy Sciences, under Contract No.~DE-SC0020330.
\end{acknowledgments}

\appendix

\section{Crank-Nicolson method}\label{sec:cranknicolson}

In this Appendix, we describe, for pedagogical purposes, the numerical implementation used to solve the nonlinear stochastic FitzHugh-Nagumo reaction-diffusion equations, Eq.~(\ref{eq:fhn}), driven by the temporally correlated noise $\eta(\ell,t)$ characterized by Eq.~(\ref{eq:noiseproperties}). An equivalent description of the correlated noise is provided by the stochastic differential equation Eq.~(\ref{eq:noiseeq}), in which $\eta$ evolves under a Langevin white noise $\xi$. For the numerical simulations, Eq.~\eqref{eq:noiseeq} is discretized using the Euler-Maruyama scheme,
\begin{equation}
\label{eq:noise_discrete}
    \eta_i(t+\Delta t)
    =\left(1-\frac{\Delta t}{\tau_c}\right)\eta_i(t)
    +\frac{\sqrt{2A\Delta t}}{\tau_c}\,\xi_i(t),
\end{equation}
where $i$ labels the spatial grid points and each $\xi_i(t)$ is an independent standard Gaussian random variable. The factor $\sqrt{\Delta t}$ ensures the correct continuum-time scaling. At each time step, the updated noise field $\eta_i(t)$ enters the activator dynamics through Eq.~\eqref{eq:fhn_u}.

The FHN equations are advanced in time using the Crank-Nicolson method, an implicit scheme that is second-order accurate in time. We adopt this method because it is unconditionally stable for conventional \emph{linear} diffusion equations, whereas explicit methods require substantially smaller time steps to maintain stability at the same spatial resolution. The principal complication arises from the nonlinear reaction term in the activator equation, Eq.~\eqref{eq:fhn_u}, which we write as
\begin{align}
    \varphi(\ell,t)
    \equiv
    \frac{1}{\tau_u}(1-u^2)(u-v).
\end{align}
The nonlinear dependence of $\varphi$ on $u$ and $v$ requires the solution of a nonlinear system at every time step. Although one could simplify the problem by linearizing $\varphi$, such an approximation is only appropriate when the fluctuations of $u$ and $v$ remain small. Instead, we retain the full nonlinearity and determine the solution using a fixed-point iteration. Consequently, the time step must be chosen sufficiently small to resolve the reaction dynamics, ensure convergence of the nonlinear iteration, and accurately integrate the stochastic noise process.

The periodic spatial domain is discretized into \(N\) uniformly spaced grid points,
\begin{align}
\ell_i=ia,~~~a=L/N,
\end{align}
for $i=0,\ldots,N-1$, with all spatial indices understood modulo $N$. Let $\mathbf L$ denote the standard second-order finite-difference Laplacian with periodic boundary conditions. At time $t_n=n\Delta t$, the discretized fields are denoted by $\boldsymbol u^n$ and $\boldsymbol v^n$, and the nonlinear reaction term is evaluated componentwise as
\begin{align}
\varphi_i^n
=
\frac{1}{\tau_u}
\left[1-(u_i^n)^2\right]
\left(u_i^n-v_i^n\right).
\end{align}

Applying the Crank-Nicolson discretization to the activator and inhibitor equations yields
\begin{align}
\frac{\boldsymbol u^{n+1}-\boldsymbol u^n}{\Delta t}
&=
\frac{D_u}{2}\mathbf L
\left(\boldsymbol u^{n+1}+\boldsymbol u^n\right)
+
\frac{\boldsymbol\varphi^{n+1}+\boldsymbol\varphi^n}{2},
\label{eq:cn-u}\\
\frac{\boldsymbol v^{n+1}-\boldsymbol v^n}{\Delta t}
&=
\frac{D_v}{2}\mathbf L
\left(\boldsymbol v^{n+1}+\boldsymbol v^n\right)
\notag\\
+\frac{1}{2\tau_v}&
\left[
\boldsymbol u^{n+1}+\boldsymbol u^n
-\alpha\left(\boldsymbol v^{n+1}+\boldsymbol v^n\right)
+2\beta\boldsymbol 1
\right],
\label{eq:cn-v}
\end{align}
where $\boldsymbol 1$ denotes the constant unit vector. Introducing the coefficients
\begin{align}
\label{eq:cn-coefficients}
\lambda_u&=\frac{D_u\Delta t}{2},&
\lambda_v&=\frac{D_v\Delta t}{2},&
c_v&=\frac{\Delta t}{2\tau_v},
\end{align}
the corresponding system matrices become
\begin{align}
\mathbf A_u&=\mathbf I-\lambda_u\mathbf L,&
\mathbf A_v&=\mathbf I-\lambda_v\mathbf L+\alpha c_v\mathbf I,\\
\mathbf B_u&=\mathbf I+\lambda_u\mathbf L,&
\mathbf B_v&=\mathbf I+\lambda_v\mathbf L-\alpha c_v\mathbf I.
\end{align}
The coupled equations can then be written compactly as
\begin{align}
&\begin{bmatrix}
\mathbf A_u & \mathbf 0\\
-c_v\mathbf I & \mathbf A_v
\end{bmatrix}
\begin{bmatrix}
\boldsymbol u^{n+1}\\
\boldsymbol v^{n+1}
\end{bmatrix}
\notag\\
&\qquad=
\begin{bmatrix}
\mathbf B_u & \mathbf 0\\
c_v\mathbf I & \mathbf B_v
\end{bmatrix}
\begin{bmatrix}
\boldsymbol u^n\\
\boldsymbol v^n
\end{bmatrix}
+
\begin{bmatrix}
\dfrac{\Delta t}{2}
\left(
\boldsymbol\varphi^n+
\boldsymbol\varphi^{n+1}
\right)\\
2\beta c_v\boldsymbol 1
\end{bmatrix}.
\label{eq:cn-block-system}
\end{align}
The coefficient matrices on the left-hand side, as well as the matrix multiplying the previous time step solution on the right-hand side, are independent of time and therefore need to be constructed only once at the beginning of the simulation.

The remaining challenge is that $\boldsymbol\varphi^{n+1}$ depends nonlinearly on the unknown fields $\boldsymbol u^{n+1}$ and $\boldsymbol v^{n+1}$. We therefore solve Eq.~\eqref{eq:cn-block-system} using a fixed-point iteration. The iteration is initialized with
\begin{align}
\boldsymbol\varphi^{n+1,0}
=
\boldsymbol\varphi^n,
\end{align}
and, given the nonlinear term from the $k$th iteration, the updated fields are obtained from
\begin{align}
&\begin{bmatrix}
\mathbf A_u & \mathbf 0\\
-c_v\mathbf I & \mathbf A_v
\end{bmatrix}
\begin{bmatrix}
\boldsymbol u^{n+1,k+1}\\
\boldsymbol v^{n+1,k+1}
\end{bmatrix}
\notag\\
&\qquad=
\begin{bmatrix}
\mathbf B_u & \mathbf 0\\
c_v\mathbf I & \mathbf B_v
\end{bmatrix}
\begin{bmatrix}
\boldsymbol u^n\\
\boldsymbol v^n
\end{bmatrix}
+
\begin{bmatrix}
\dfrac{\Delta t}{2}
\left(
\boldsymbol\varphi^n+
\boldsymbol\varphi^{n+1,k}
\right)\\
2\beta c_v\boldsymbol 1
\end{bmatrix}.
\label{eq:cn-fixed-point-system}
\end{align}
The nonlinear reaction term is then updated according to
\begin{align}
\varphi_i^{n+1,k+1}
=
\frac{1}{\tau_u}
\left[
1-
\left(
u_i^{n+1,k+1}
\right)^2
\right]
\left(
u_i^{n+1,k+1}
-
v_i^{n+1,k+1}
\right),
\end{align}
and the iteration is repeated until the residual
\begin{align}
\left\|
\boldsymbol\varphi^{n+1,k+1}
-
\boldsymbol\varphi^{n+1,k}
\right\|
\end{align}
falls below a prescribed tolerance, or until a specified maximum number of iterations is reached. Once the solution at the new time step has converged, the stochastic forcing is incorporated by updating the activator field using the Euler-Maruyama noise increment given in Eq.~\eqref{eq:noise_discrete}.

\section{Nonequilibrium quiescent background}\label{sec:noneq}

In the presence of Gaussian white noise, we find that the mean values of the activator and inhibitor values are shifted compared to their quiescent values in the absence of noise, $u_0 = -1$ and $v_0 = ( \beta-1)/\alpha$. A nonequilibrium quiescent background emerges, with activator and inhibitor values now fluctuating near new nonequilibrium quiescent means $\mu_u$ and $\mu_v$, respectively. Fluctuations must excite $u$ above the new nonequilibrium inhibitor mean $\mu_v$ for the nonlinear amplification process to take over and nucleate a localized pattern. In this Appendix, we discuss how to analytically determine the nonequilibrium quiescent means, and predict if the integrated noise strength $A$ is sufficiently strong for the system to enter the pattern-forming regime.

While the full FitzHugh-Nagumo equations cannot be fully analytically solved, applying some simple approximations allows us to analytically characterize the behavior of the nonequilibrium quiescent state. These approximations are motivated by observations of the simulations. Noting that the noise is spatially homogeneous, we perform spatial averaging over the periodic 1D system, where henceforth we use $\overline X \equiv L^{-1}\oint \t d\ell\,X(\ell)$. Applying the averaging to both sides of the reaction-diffusion equation subject to Gaussian white noise, we find
\begin{align}
    \frac{\partial \overline u}{\partial t} &= \frac{1}{\tau_u} \left(\overline u  - \overline v  - \overline {u^3}+ \overline{u^2 v}  \right)\\
    \frac{\partial \overline v}{\partial t} &= \frac{1}{\tau_v}(\overline u  - \alpha \overline v + \beta)
    \end{align}
Here, the diffusive terms are eliminated because the averaging is done over a spatially periodic system and $\overline \eta  \approx 0$ because the noise has zero mean.

In the nonequilibrium quiescent background, the spatial averages $\overline u$ and $\overline v$ are time-independent, and are denoted by $\mu_u$ and $\mu_v$, respectively. Moreover, in the absence of noise they revert to $\mu_u = u_0$ and $\mu_v = v_0$. Approximating that $u(\ell)$ is spatially Gaussian distributed and neglecting the covariance between $u^2$ and $v$, the equations of motion reduce to
\begin{subequations}\label{eq:sseom}
\begin{align}
    0 &\approx \mu_u - \mu_v - \mu_u^3 - 3\mu_u\sigma_u^2 + (\mu_u^2 + \sigma_u^2)\mu_v,\label{eq:sseom1}\\
    0 &\approx \mu_u - \alpha \mu_v + \beta\label{eq:sseom2}
\end{align}
\end{subequations}
for the nonequilibrium quiescent background. The coupled equations above relate the nonequilibrium quiescent background activator and inhibitor averages, $\mu_u$ and $\mu_v$, to $\sigma_u^2$. 

The variance $\sigma_u^2$ can be estimated in a straightforward linearized treatment. To this end, we first begin by expanding the values of $u(\ell)$ and $v(\ell)$ around $\mu_u$ and $\mu_v$,
\begin{align}
    u(\ell) = \mu_u + \delta u(\ell),~~~v(\ell) = \mu_v + \delta v(\ell).
\end{align}
The latter will be solved for self-consistently, upon including Eq.~\eqref{eq:sseom}. Since $\mu_u$ is equated to the steady-state spatial average of $u$, the spatial average of the fluctuations $\delta u$ is zero. Moreover, through empirical observation of simulations of the nonequilibrium quiescent background, we find that the fluctuations $\delta v/\mu_v\ll 1$ are small. Hence, when we linearize the equation of motion for activator $u$, we can neglect contributions proportional to $\delta v$, yielding
\begin{align}
    \partial_t \delta u &= D_u\partial_\ell^2 \delta u +\tau_u^{-1}(1- 3\mu_u^2 +2\mu_u\mu_v)\delta u + \eta.
\end{align}
Applying a discrete spatial Fourier transform, we can make the substitution $ \partial_\ell^2\delta u \rightarrow - 2a^{-2}\left(1- \cos\frac{2\pi m}{N}\right)\delta u_m$, where $\delta u_m$ is the amplitude of each Fourier mode. The governing linearized equation for the modes becomes
\begin{subequations}
\begin{align}
    \partial_t \delta u_m =  -r_m \delta u_m + \eta_m,
\end{align}
where
\begin{align}
    r_m= \frac{3\mu_u^2 - 2\mu_u\mu_v-1}{\tau_u} + \frac{2D_u}{a^2}\left(1- \cos\frac{2\pi m}{N}\right)
\end{align}
\end{subequations}
is the damping coefficient for Fourier mode $m$. Examining the damping coefficient, we see that the first term suppresses long wavelength fluctuations over the time scale of $\tau_u$ while short wavelength fluctuations are suppressed over the diffusion time scale $a^2/D_u$.

Time discretization is performed by using the Crank-Nicolson method. Doing so, we arrive at the implicit equation
\begin{align}\label{supeq:impliciteq}
    \delta u_m^{n+1} = \frac{1 - r_m\Delta t/2}{1+r_m\Delta t/2}\delta u _m^{n} + \sqrt{2A\Delta t}\, \xi_m^n,
\end{align}
where $\xi_m^n$ is the Gaussian white noise sample driving mode $m$ at time step $n$ and $A$ is the integrated noise strength defined in Eq.~\eqref{eq:integratednoise}. We expect that the fluctuations $\delta u_m^n$ are independent of the time step $n$.  These statistical properties can be quantified by the \textit{ensemble variance} of the mode $m$ via $\langle|\delta u_m^n|^2\rangle$, where the averaging $\langle \cdots \rangle$ is done over an ensemble of simulations at time step $n$. Hence, we expect that for self-consistency, $\langle |\delta u_m^n|^2\rangle = \langle|\delta u_m^{n'}|^2\rangle$ for all values of $n$ and $n'$. 

To find the ensemble variance for the mode $m$, we square both sides of Eq.~\eqref{supeq:impliciteq} and perform the ensemble average. We apply the stationary condition $\langle |\delta u_m^n|^2\rangle = \langle|\delta u_m^{n+1}|^2\rangle$. Finally, we arrive at the expression for the stationary ensemble variance of mode $m$,
\begin{align}\label{supeq:stationaryvariance}
    \left\langle |\delta u_m|^2\right\rangle = \frac{A}{r_m}\left(1+\frac{r_m \Delta t}{2}\right)^2.
\end{align}
Here, we have also used $\left\langle |\xi^n_m|^2\right\rangle = 1$ and that $\delta u_m^n$ and $\xi^n_m$ are independent, so have zero covariance. At this stage, we can apply Parseval's identity to relate the spatial variance of $\delta u$ to the Fourier modes
\begin{equation}
    \sigma_u^2 = \overline {(\delta u)^2} = \frac{1}{N}\sum_{m=1}^{N-1}(\delta u_m)^2.
\end{equation}
Taking the ensemble average of both sides and substituting in Eq.~\eqref{supeq:stationaryvariance} for each term in the sum in the right-hand side, we can approximate $\sigma_u^2$ as its ensemble average, yielding
\begin{align}\label{supeq:variancefinal}
    \sigma_u^2 \approx  \left \langle \overline{(\delta u )^2}\right \rangle  = \frac{1}{N}\sum_{m=1}^{N-1}\frac{A}{r_m}\left(1+\frac{r_m \Delta t}{2}\right)^2.
\end{align}
Solving the system of equations of Eqs.~\eqref{eq:sseom} and \eqref{supeq:variancefinal} yields estimates for the values of  $\mu_u$, $\mu_v$, and $\sigma_u^2$. While the approximations taken to arrive at these results were rather crude, the estimates still retain the key qualitative features and have excellent numerical agreement with the simulated results. Equation~\eqref{supeq:variancefinal} also reveals the dependence of the spread of the activator on the discretization time $\Delta t$. For the nonequilibrium-state to agree irrespective of the chosen $\Delta t$, a sufficient condition is that $D_u\Delta t/a^2 \ll 1$.

In the continuous time-limit of $D_u\Delta t/a^2\rightarrow 0$, we find that the variance simplifies to 
\begin{align}\label{eq:variance}
    \sigma_u^2 \approx \frac{A}{\sqrt{\kappa(\kappa +4D_u/a^2)}},~~~\kappa = \frac{3\mu_u^2 - 2\mu_u\mu_v-1}{\tau_u}.
\end{align}
Here, $\kappa$ can be understood as the restoring rate of fluctuations of $u$ away from $\mu_u$. We have also evaluated these quantities in the limit of $N\rightarrow\infty$. Putting things all together, self-consistently solving the system of equations comprised of Eqs.~\eqref{eq:sseom1}, \eqref{eq:sseom2}, and \eqref{eq:variance} for $\mu_u$, $\mu_v$, and $\sigma_u$ fully characterizes the nonequilibrium quiescent background.

From Eq.~\eqref{eq:variance}, we see that $\sigma_u^2\propto A$, meaning that in the absence of noise $\sigma_u = 0$ and $\mu_u$ and $\mu_v$ revert to their quiescent values of $u_0$ and $v_0$. Gaussian white noise introduces a finite $\sigma_u^2$, which shifts $\mu_u$ and $\mu_v$ from the quiescent values. For Gaussian white noise to trigger pattern formation, we expect that the integrated noise strength $A$ is sufficiently large such that the spread comparable to the difference between the quiescent values of $u_0$ and $v_0$, namely $\sigma_u \sim (\alpha + \beta - 1)/\alpha$.

\begin{figure*}
    \centering
    \includegraphics[width=0.9\linewidth]{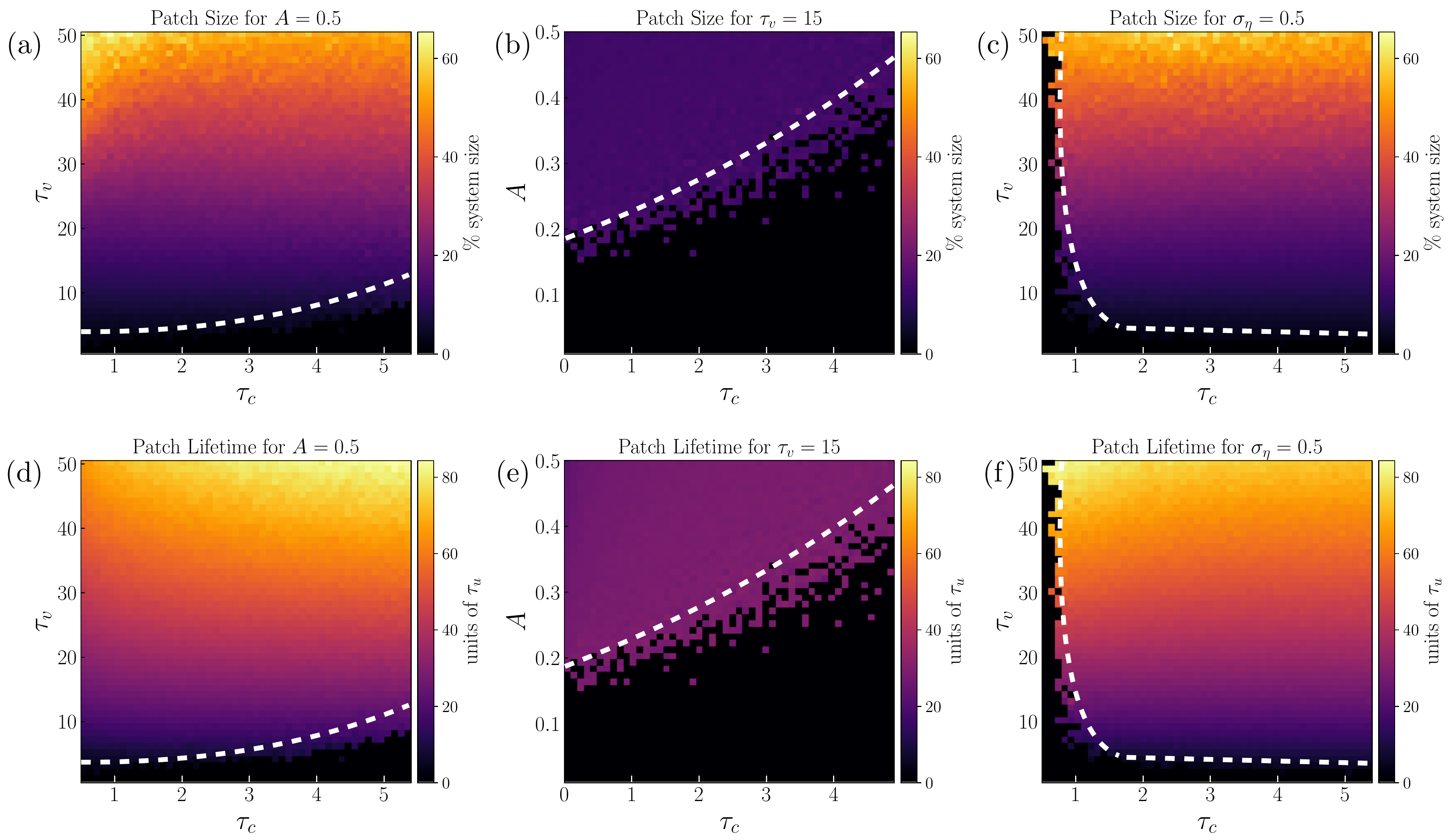}
    \caption{Properties of the noise-induced excitation patches for the three parameter sweeps corresponding to Fig.~\ref{fig:phase-diagram}. Panels (a)--(c) show the characteristic patch size as a percentage of the system size, while panels (d)--(f) show the characteristic patch lifetime in units of $\tau_u$. Panels (a) and (d) vary $\tau_v$ and $\tau_c$ at fixed integrated noise strength $A=0.5$; panels (b) and (e) vary $A$ and $\tau_c$ at fixed inhibitor response time $\tau_v=15$; and panels (c) and (f) vary $\tau_v$ and $\tau_c$ at fixed equal-time rms noise amplitude $\sigma_\eta=0.5$. The white dashed curves reproduce the dynamical phase boundaries from Fig.~\ref{fig:phase-diagram}, and black regions indicate parameter combinations for which no excitation patches were detected. The remaining parameters are $\alpha=0.5$, $\beta=0.7$, $D_u=1$, and $D_v=5$.}
    \label{fig:phase-diagram_tv_vs_tc_fixed_sigma_eta}
\end{figure*}

\section{Properties of excited patches}\label{section:properties}

In this Appendix, we examine how the characteristic size and lifetime of the excitation patches vary across the parameter ranges explored in the three phase diagrams of Fig.~\ref{fig:phase-diagram}. Figure~\ref{fig:phase-diagram_tv_vs_tc_fixed_sigma_eta} shows the patch size (top row) and lifetime (bottom row) for the same three parameter sweeps. In the first and third columns, $\tau_v$ and $\tau_c$ are varied at fixed integrated noise strength $A$ and fixed equal-time rms amplitude $\sigma_\eta$, respectively, while the middle column varies $A$ and $\tau_c$ at fixed $\tau_v$.

A common feature of these results is that, once a patch is nucleated, its size and lifetime are governed predominantly by the inhibitor response time $\tau_v$. Both quantities increase approximately linearly with $\tau_v$ over much of the pattern-forming regime, consistent with Fig.~\ref{fig:size-lifetime}. This agrees with the physical picture that a slower inhibitor allows an excitation to persist longer and spread farther before inhibitory feedback terminates its growth. In particular, when $\tau_c<\tau_v$, the correlation time strongly affects whether pattern formation occurs, as reflected by the phase boundaries, but has comparatively little effect on the size and lifetime of the resulting patches.

The three noise parameterizations further clarify the distinct roles of the stochastic forcing and inhibitor dynamics. At fixed $A$, increasing $\tau_c$ shifts the nucleation boundary because increased persistence is accompanied by reduced instantaneous noise amplitude, whereas at fixed $\sigma_\eta$, increasing $\tau_c$ strongly promotes nucleation because persistence increases without an amplitude penalty. Despite these opposite effects on the onset of pattern formation, in both cases the size and lifetime of the resulting patches depend predominantly on $\tau_v$. The middle column provides a complementary perspective by fixing $\tau_v$ while varying $A$ and $\tau_c$. Here, both noise parameters strongly affect the boundary for patch nucleation, while the characteristic size and lifetime of the patches show a comparatively weaker dependence within the pattern-forming regime.

A different behavior emerges when the noise persists on a timescale comparable to or longer than the inhibitor response time, $\tau_c\gtrsim\tau_v$. This effect is most clearly visible in the patch-lifetime map at fixed $\sigma_\eta$, Fig.~\ref{fig:phase-diagram_tv_vs_tc_fixed_sigma_eta}(f), where the lifetime develops an appreciable dependence on $\tau_c$ in the regime of relatively small $\tau_v$ and large $\tau_c$. In this regime, a sufficiently strong and long-lived fluctuation can continue to sustain the activator even after the inhibitor has risen to suppress it, thereby prolonging the excitation. Consequently, for $\tau_c\gg\tau_v$, the patch lifetime need no longer be determined primarily by $\tau_v$ and can instead become controlled by the persistence time $\tau_c$. Thus, while the noise primarily controls nucleation for $\tau_c<\tau_v$, sufficiently persistent noise can also directly influence the subsequent dynamics of an already nucleated patch.

\bibliography{ref.bib}

@article{fitzhugh61,
  author  = {FitzHugh, Richard},
  title   = {Impulses and Physiological States in Theoretical Models of Nerve Membrane},
  journal = {Biophys. J.},
  volume  = {1},
  number  = {6},
  pages   = {445--466},
  year    = {1961},
  url = {https://doi.org/10.1016/S0006-3495(61)86902-6}
}

@article{nagumo62,
  author  = {Nagumo, {Jin-Ichi} and Arimoto, Suguru and Yoshizawa, Shuji},
  title   = {An Active Pulse Transmission Line Simulating Nerve Axon},
  journal = {Proc. IEEE},
  volume  = {50},
  number  = {10},
  pages   = {2061--2070},
  year    = {1962},
  url = {https://ieeexplore.ieee.org/document/4066548}
}

@article{vicker00,
title = {Reaction–diffusion waves of actin filament polymerization/depolymerization in Dictyostelium pseudopodium extension and cell locomotion},
journal = {Biophys. Chem.},
volume = {84},
number = {2},
pages = {87-98},
year = {2000},
issn = {0301-4622},
url = {https://www.sciencedirect.com/science/article/pii/S0301462299001465},
author = {Michael G. Vicker}
}

@article{gunther04,
title = {Mobile Actin Clusters and Traveling Waves in Cells Recovering from Actin Depolymerization},
journal = {Biophys. J.},
volume = {87},
number = {5},
pages = {3493-3503},
year = {2004},
issn = {0006-3495},
url = {https://www.sciencedirect.com/science/article/pii/S000634950473814X},
author = {Günther Gerisch and Till Bretschneider and Annette Müller-Taubenberger and Evelyn Simmeth and Mary Ecke and Stefan Diez and Kurt Anderson}
}

@article{weiner07,
    author = {Weiner, Orion D AND Marganski, William A AND Wu, Lani F AND Altschuler, Steven J AND Kirschner, Marc W},
    journal = {PLOS Biol.},
    publisher = {Public Library of Science},
    title = {An Actin-Based Wave Generator Organizes Cell Motility},
    year = {2007},
    month = {08},
    volume = {5},
    url = {https://doi.org/10.1371/journal.pbio.0050221},
    pages = {1-11},
    number = {9},

}

@article{asano08,
author = {Asano, Yukako and Nagasaki, Akira and Uyeda, Taro Q.P.},
title = {Correlated waves of actin filaments and PIP3 in Dictyostelium cells},
journal = {Cell Motil.},
volume = {65},
number = {12},
pages = {923-934},
url = {https://onlinelibrary.wiley.com/doi/abs/10.1002/cm.20314},
year = {2008}
}

@Article{huang13,
author={Huang, Chuan-Hsiang
and Tang, Ming
and Shi, Changji
and Iglesias, Pablo A.
and Devreotes, Peter N.},
title={An excitable signal integrator couples to an idling cytoskeletal oscillator to drive cell migration},
journal={Nat. Cell Biol.},
year={2013},
month={Nov},
day={01},
volume={15},
number={11},
pages={1307-1316},
issn={1476-4679},
url={https://doi.org/10.1038/ncb2859}
}

@article{arai10,
author = {Yoshiyuki Arai  and Tatsuo Shibata  and Satomi Matsuoka  and Masayuki J. Sato  and Toshio Yanagida  and Masahiro Ueda },
title = {Self-organization of the phosphatidylinositol lipids signaling system for random cell migration},
journal = {Proc. Natl. Acad. Sci. U.S.A.},
volume = {107},
number = {27},
pages = {12399-12404},
year = {2010},
URL = {https://www.pnas.org/doi/abs/10.1073/pnas.0908278107}}

@article{allard13,
title = {Traveling waves in actin dynamics and cell motility},
journal = {Curr. Opin. Cell Biol.},
volume = {25},
number = {1},
pages = {107-115},
year = {2013},
issn = {0955-0674},
url = {https://www.sciencedirect.com/science/article/pii/S0955067412001378},
author = {Jun Allard and Alex Mogilner}
}

@article{taniguchi13,
author = {Daisuke Taniguchi  and Shuji Ishihara  and Takehiko Oonuki  and Mai Honda-Kitahara  and Kunihiko Kaneko  and Satoshi Sawai },
title = {Phase geometries of two-dimensional excitable waves govern self-organized morphodynamics of amoeboid cells},
journal = {Proc. Natl. Acad. Sci. U.S.A.},
volume = {110},
number = {13},
pages = {5016-5021},
year = {2013},
URL = {https://www.pnas.org/doi/abs/10.1073/pnas.1218025110}}

@article{xiong10,
author = {Yuan Xiong  and Chuan-Hsiang Huang  and Pablo A. Iglesias  and Peter N. Devreotes },
title = {Cells navigate with a local-excitation, global-inhibition-biased excitable network},
journal = {Proc. Natl. Acad. Sci. U.S.A.},
volume = {107},
number = {40},
pages = {17079-17086},
year = {2010},
URL = {https://www.pnas.org/doi/abs/10.1073/pnas.1011271107}}

@article{nishikawa14,
title = {Excitable Signal Transduction Induces Both Spontaneous and Directional Cell Asymmetries in the Phosphatidylinositol Lipid Signaling System for Eukaryotic Chemotaxis},
journal = {Biophys. J.},
volume = {106},
number = {3},
pages = {723-734},
year = {2014},
issn = {0006-3495},
url = {https://www.sciencedirect.com/science/article/pii/S0006349513058153},
author = {Masatoshi Nishikawa and Marcel Hörning and Masahiro Ueda and Tatsuo Shibata}
}

@article{haastert17,
author = {van Haastert, Peter J. M. and Keizer-Gunnink, Ineke and Kortholt, Arjan},
title = {Coupled excitable Ras and F-actin activation mediates spontaneous pseudopod formation and directed cell movement},
journal = {Mol. Biol. Cell},
volume = {28},
number = {7},
pages = {922-934},
year = {2017},
URL = {https://doi.org/10.1091/mbc.e16-10-0733},
}

@article{devreotes17,
   author = "Devreotes, Peter N. and Bhattacharya, Sayak and Edwards, Marc and Iglesias, Pablo A. and Lampert, Thomas and Miao, Yuchuan",
   title = "Excitable Signal Transduction Networks in Directed Cell Migration", 
   journal= "Annu. Rev. Cell Dev. Biol.",
   year = "2017",
   volume = "33",
   number = "Volume 33, 2017",
   pages = "103-125",
   url = "https://www.annualreviews.org/content/journals/10.1146/annurev-cellbio-100616-060739",
   publisher = "Annual Reviews",
   issn = "1530-8995",
   type = "Journal Article",
  }

@ARTICLE{matsuoka24,
AUTHOR={Matsuoka, Satomi and Iwamoto, Koji and Shin, Da Young and Ueda, Masahiro},   
TITLE={Spontaneous signal generation by an excitable system for cell migration},      
JOURNAL={Front. Cell Dev. Biol.},      
VOLUME={12},           
YEAR={2024},      
URL={https://www.frontiersin.org/articles/10.3389/fcell.2024.1373609},
ISSN={2296-634X},   
}

@article{zhan20,
title = {An Excitable Ras/PI3K/ERK Signaling Network Controls Migration and Oncogenic Transformation in Epithelial Cells},
journal = {Dev. Cell.},
volume = {54},
number = {5},
pages = {608-623.e5},
year = {2020},
issn = {1534-5807},
url = {https://www.sciencedirect.com/science/article/pii/S1534580720305992},
author = {Huiwang Zhan and Sayak Bhattacharya and Huaqing Cai and Pablo A. Iglesias and Chuan-Hsiang Huang and Peter N. Devreotes},
}

@Article{zhan25,
author={Zhan, Huiwang
and Pal, Dhiman Sankar
and Borleis, Jane
and Deng, Yu
and Long, Yu
and Janetopoulos, Chris
and Huang, Chuan-Hsiang
and Devreotes, Peter N.},
title={Self-organizing glycolytic waves tune cellular metabolic states and fuel cancer progression},
journal={Nat. Commun.},
year={2025},
month={Jul},
day={01},
volume={16},
number={1},
pages={5563},
issn={2041-1723},
url={https://doi.org/10.1038/s41467-025-60596-6}
}

@article{aoki13,
title = {Stochastic ERK Activation Induced by Noise and Cell-to-Cell Propagation Regulates Cell Density-Dependent Proliferation},
journal = {Mol. Cell.},
volume = {52},
number = {4},
pages = {529-540},
year = {2013},
issn = {1097-2765},
url = {https://www.sciencedirect.com/science/article/pii/S1097276513006850},
author = {Kazuhiro Aoki and Yuka Kumagai and Atsuro Sakurai and Naoki Komatsu and Yoshihisa Fujita and Clara Shionyu and Michiyuki Matsuda},
}

@article{albeck13,
title = {Frequency-Modulated Pulses of ERK Activity Transmit Quantitative Proliferation Signals},
journal = {Mol. Cell.},
volume = {49},
number = {2},
pages = {249-261},
year = {2013},
issn = {1097-2765},
url = {https://www.sciencedirect.com/science/article/pii/S1097276512009331},
author = {John G. Albeck and Gordon B. Mills and Joan S. Brugge},
}

@Article{yang18,
author={Yang, Jr-Ming
and Bhattacharya, Sayak
and West-Foyle, Hoku
and Hung, Chien-Fu
and Wu, T.-C.
and Iglesias, Pablo A.
and Huang, Chuan-Hsiang},
title={Integrating chemical and mechanical signals through dynamic coupling between cellular protrusions and pulsed ERK activation},
journal={Nat. Commun.},
year={2018},
month={Nov},
day={07},
volume={9},
number={1},
pages={4673},
issn={2041-1723},
url={https://doi.org/10.1038/s41467-018-07150-9}
}

@article{vega18,
title = {Oncogenic Signaling Pathways in The Cancer Genome Atlas},
journal = {Cell},
volume = {173},
number = {2},
pages = {321-337.e10},
year = {2018},
issn = {0092-8674},
url = {https://www.sciencedirect.com/science/article/pii/S0092867418303593},
author = {F. Sanchez-Vega and Marco Mina and Joshua Armenia and Walid K. Chatila and Augustin Luna and Konnor C. La and others},
}

@article{watton99,
title = {Akt/PKB localisation and 3' phosphoinositide generation at sites of epithelial cell-matrix and cell-cell interaction},
journal = {Curr. Biol.},
volume = {9},
number = {8},
pages = {433-436},
year = {1999},
issn = {0960-9822},
url = {https://www.sciencedirect.com/science/article/pii/S0960982299801924},
author = {Sandra J. Watton and Julian Downward},
}

@article{hecht2010,
  title = {Transient Localized Patterns in Noise-Driven Reaction-Diffusion Systems},
  author = {Hecht, Inbal and Kessler, David A. and Levine, Herbert},
  journal = {Phys. Rev. Lett.},
  volume = {104},
  issue = {15},
  pages = {158301},
  numpages = {4},
  year = {2010},
  month = {Apr},
  publisher = {American Physical Society},
  url = {https://link.aps.org/doi/10.1103/PhysRevLett.104.158301}
}

@article{hechtPCB11,
  author = {Hecht, Inbal and Skoge, Monica L and Charest, Pascale G and Ben-Jacob, Eshel and Firtel, Richard A and Loomis, William F and Levine, Herbert and Rappel, Wouter-Jan},
  title = {Activated membrane patches guide chemotactic cell motility},
  journal = {PLoS Comput. Biol.},
  year = {2011},
  volume = {7},
  number = {6},
  pages = {e1002044},
  url = {https://doi.org/10.1371/journal.pcbi.1002044},
  issn = {1553-7358}
}

@incollection{hanggi95,
  author    = {H{\"a}nggi, Peter and Jung, Peter},
  title     = {Colored Noise in Dynamical Systems},
  booktitle = {Adv. Chem. Phys.},
  volume    = {89},
  pages     = {239--326},
  year      = {1995},
  publisher = {John Wiley \& Sons},
  url       = {https://doi.org/10.1002/9780470141489.ch4}
}

@article{ramaswamy10,
  author  = {Ramaswamy, Sriram},
  title   = {The Mechanics and Statistics of Active Matter},
  journal = {Annu. Rev. Condens. Matter Phys.},
  volume  = {1},
  pages   = {323--345},
  year    = {2010},
  url     = {https://doi.org/10.1146/annurev-conmatphys-070909-104101}
}

@article{marchetti13,
  author  = {Marchetti, M. C. and Joanny, J.-F. and Ramaswamy, S. and
             Liverpool, T. B. and Prost, J. and Rao, M. and Simha, R. A.},
  title   = {Hydrodynamics of Soft Active Matter},
  journal = {Rev. Mod. Phys.},
  volume  = {85},
  pages   = {1143--1189},
  year    = {2013},
  url     = {https://doi.org/10.1103/RevModPhys.85.1143}
}

@article{elgeti15,
  author  = {Elgeti, Jens and Winkler, Roland G. and Gompper, Gerhard},
  title   = {Physics of Microswimmers---Single Particle Motion and
             Collective Behavior: A Review},
  journal = {Rep. Prog. Phys.},
  volume  = {78},
  number  = {5},
  pages   = {056601},
  year    = {2015},
  url     = {https://doi.org/10.1088/0034-4885/78/5/056601}
}

@article{cates15,
  author  = {Cates, Michael E. and Tailleur, Julien},
  title   = {Motility-Induced Phase Separation},
  journal = {Annu. Rev. Condens. Matter Phys.},
  volume  = {6},
  pages   = {219--244},
  year    = {2015},
  url     = {https://doi.org/10.1146/annurev-conmatphys-031214-014710}
}

@article{bechinger16,
  author  = {Bechinger, Clemens and Di Leonardo, Roberto and L{\"o}wen, Hartmut
             and Reichhardt, Charles and Volpe, Giorgio and Volpe, Giovanni},
  title   = {Active Particles in Complex and Crowded Environments},
  journal = {Rev. Mod. Phys.},
  volume  = {88},
  pages   = {045006},
  year    = {2016},
  url     = {https://doi.org/10.1103/RevModPhys.88.045006}
}

@article{tevrugt26,
  author  = {te Vrugt, Michael and Liebchen, Benno and Cates, Michael E.},
  title   = {Colloquium: What Do We Mean by ``Active Matter''?},
  journal = {Rev. Mod. Phys.},
  volume  = {98},
  pages   = {031001},
  year    = {2026},
  url     = {https://doi.org/10.1103/wd4f-q7kv}
}

@article{adamer20,
  author  = {Adamer, Michael F. and Harrington, Heather A. and
             Gaffney, Eamonn A. and Woolley, Thomas E.},
  title   = {Coloured Noise from Stochastic Inflows in Reaction--Diffusion Systems},
  journal = {Bull. Math. Biol.},
  volume  = {82},
  pages   = {44},
  year    = {2020},
  url = {https://link.springer.com/article/10.1007/s11538-020-00719-w}
}

@article{kaern05,
  author  = {K{\ae}rn, Mads and Elston, Timothy C. and Blake, William J.
             and Collins, James J.},
  title   = {Stochasticity in Gene Expression: From Theories to Phenotypes},
  journal = {Nat. Rev. Genet.},
  volume  = {6},
  pages   = {451--464},
  year    = {2005},
  url     = {https://doi.org/10.1038/nrg1615}
}

@article{raser05,
  author  = {Raser, Jonathan M. and O'Shea, Erin K.},
  title   = {Noise in Gene Expression: Origins, Consequences, and Control},
  journal = {Science},
  volume  = {309},
  number  = {5743},
  pages   = {2010--2013},
  year    = {2005},
  url     = {https://doi.org/10.1126/science.1105891}
}

@article{briat23,
  author  = {Briat, Corentin and Khammash, Mustafa},
  title   = {Noise in Biomolecular Systems: Modeling, Analysis, and
             Control Implications},
  journal = {Annu. Rev. Control Robot. Auton. Syst.},
  volume  = {6},
  pages   = {283--311},
  year    = {2023},
  url     = {https://doi.org/10.1146/annurev-control-042920-101825}
}

@article{martin2021,
  title = {Statistical mechanics of active Ornstein-Uhlenbeck particles},
  author = {Martin, David and O'Byrne, J\'er\'emy and Cates, Michael E. and Fodor, \'Etienne and Nardini, Cesare and Tailleur, Julien and van Wijland, Fr\'ed\'eric},
  journal = {Phys. Rev. E},
  volume = {103},
  issue = {3},
  pages = {032607},
  numpages = {25},
  year = {2021},
  month = {Mar},
  publisher = {American Physical Society},
  url = {https://link.aps.org/doi/10.1103/PhysRevE.103.032607}
}

@article{brugioni05,
  author  = {Brugioni, S. and Hwang, D.-U. and Meucci, R. and Boccaletti, S.},
  title   = {Coherence Resonance in Excitable Electronic Circuits in the Presence of Colored Noise},
  journal = {Phys. Rev. E},
  volume  = {71},
  number  = {6},
  pages   = {062101},
  year    = {2005},
  url = {https://doi.org/10.1103/PhysRevE.71.062101}
}

@article{brandstetter10,
  author  = {Brandstetter, S. and Dahlem, M. A. and Sch{\"o}ll, E.},
  title   = {Interplay of Time-Delayed Feedback Control and Temporally Correlated Noise in Excitable Systems},
  journal = {Philos. Trans. R. Soc. A},
  volume  = {368},
  number  = {1911},
  pages   = {391--421},
  year    = {2010},
  url = {https://royalsocietypublishing.org/doi/10.1098/rsta.2009.0233}
}

@article{beato05,
  author  = {Beato, V. and Sendi{\~n}a-Nadal, I. and Gerdes, I. and Engel, H.},
  title   = {Noise-Induced Wave Nucleations in an Excitable Chemical Reaction},
  journal = {Phys. Rev. E},
  volume  = {71},
  number  = {3},
  pages   = {035204},
  year    = {2005},
  url     = {https://doi.org/10.1103/PhysRevE.71.035204}
}

@article{balanov06,
  author  = {Balanov, A. G. and Beato, V. and Janson, N. B. and Engel, H. and Sch{\"o}ll, E.},
  title   = {Delayed Feedback Control of Noise-Induced Patterns in Excitable Media},
  journal = {Phys. Rev. E},
  volume  = {74},
  number  = {1},
  pages   = {016214},
  year    = {2006},
  url     = {https://doi.org/10.1103/PhysRevE.74.016214}
}

@article{wang06,
  author  = {Wang, Hongli and Zhang, Ke and Ouyang, Qi},
  title   = {Resonant-Pattern Formation Induced by Additive Noise in
             Periodically Forced Reaction-Diffusion Systems},
  journal = {Phys. Rev. E},
  volume  = {74},
  number  = {3},
  pages   = {036210},
  year    = {2006},
  url     = {https://doi.org/10.1103/PhysRevE.74.036210}
}

@article{das13,
  author  = {Das, Debojyoti and Ray, Deb Shankar},
  title   = {Dichotomous-Noise-Induced Pattern Formation in a
             Reaction-Diffusion System},
  journal = {Phys. Rev. E},
  volume  = {87},
  number  = {6},
  pages   = {062924},
  year    = {2013},
  url     = {https://doi.org/10.1103/PhysRevE.87.062924}
}

@article{gu18,
  author  = {Gu, Anhui and Wang, Bixiang},
  title   = {Asymptotic Behavior of Random {FitzHugh--Nagumo} Systems
             Driven by Colored Noise},
  journal = {Discrete Contin. Dyn. Syst. - B},
  volume  = {23},
  number  = {4},
  pages   = {1689--1720},
  year    = {2018},
  url     = {https://doi.org/10.3934/dcdsb.2018072}
}

@article{perc05,
  author  = {Perc, Matja{\v z}},
  title   = {Spatial Coherence Resonance in Excitable Media},
  journal = {Phys. Rev. E},
  volume  = {72},
  number  = {1},
  pages   = {016207},
  year    = {2005},
  url     = {https://doi.org/10.1103/PhysRevE.72.016207}
}

@article{rinzel73,
  author  = {Rinzel, John and Keller, Joseph B.},
  title   = {Traveling Wave Solutions of a Nerve Conduction Equation},
  journal = {Biophys. J.},
  volume  = {13},
  number  = {12},
  pages   = {1313--1337},
  year    = {1973},
  url = {https://www.cell.com/biophysj/pdf/S0006-3495(73)86065-5.pdf?_returnURL=https%3A%2F%2Flinkinghub.elsevier.com%2Fretrieve%2Fpii%2FS0006349573860655%3Fshowall%3Dtrue}
}

@article{pikovskyPRL97,
  title = {Coherence Resonance in a Noise-Driven Excitable System},
  author = {Pikovsky, Arkady S. and Kurths, J\"urgen},
  journal = {Phys. Rev. Lett.},
  volume = {78},
  issue = {5},
  pages = {775--778},
  numpages = {0},
  year = {1997},
  month = {Feb},
  publisher = {American Physical Society},
  url = {https://link.aps.org/doi/10.1103/PhysRevLett.78.775}
}

@Article{miaoMSB19,
author={Miao, Yuchuan
and Bhattacharya, Sayak
and Banerjee, Tatsat
and Abubaker‐Sharif, Bedri
and Long, Yu
and Inoue, Takanari
and Iglesias, Pablo A.
and Devreotes, Peter N.},
title={Wave patterns organize cellular protrusions and control cortical dynamics},
journal={Mol. Syst. Biol.},
year={2019},
month={Mar},
day={12},
volume={15},
number={3},
pages={MSB188585},
issn={1744-4292},
url={https://doi.org/10.15252/msb.20188585}
}

@article{buschPRE03,
  title = {Influence of spatiotemporally correlated noise on structure formation in excitable media},
  author = {Busch, H. and Kaiser, F.},
  journal = {Phys. Rev. E},
  volume = {67},
  issue = {4},
  pages = {041105},
  numpages = {7},
  year = {2003},
  month = {Apr},
  publisher = {American Physical Society},
  url = {https://link.aps.org/doi/10.1103/PhysRevE.67.041105}
}

\end{document}